\documentclass[letterpaper, 12pt]{article}
\usepackage{times}
\usepackage{graphicx}
\usepackage{amsmath}
\usepackage{amssymb}
\usepackage{xspace}
\usepackage{color}
\usepackage[numbers, sort&compress,super]{natbib}

\newcommand{\ltsim}{\mbox{{\raisebox{-0.4ex}{$\stackrel{<}{{\scriptstyle\sim}}$}}}}

\def\LCMO{$\mathrm{La_{0.5}Ca_{0.5}MnO_3}$\xspace}

\title{Sliding charge density wave in manganites}

\author
{Susan Cox,$^{{1}\ast}$ J. Singleton,$^{1}$ R.D. McDonald,$^{1}$ A. Migliori,$^{1}$ P.B. Littlewood$^{2}$\\
\\
\normalsize{$^{1}$National High Magnetic Field Laboratory, Ms-E536, Los Alamos National Laboratory,}\\
\normalsize{New Mexico, 87545, USA}\\
\normalsize{$^{2}$Cavendish Laboratory, University of Cambridge, Cambridge, CB3 0HE, UK}\\
\normalsize{$^\ast$e-mail:scox@lanl.gov}
\\
}

\date{}

\begin{document} 

\baselineskip24pt

\maketitle
\newpage

The so-called stripe phase of the manganites is an important example of the complex behaviour 
of metal oxides, and has long been
interpreted as the localisation of charge at atomic sites~\cite{chen_co_melt, chen_comm_incomm, CO1, CO2a}.  
Here, we demonstrate via resistance 
measurements on La$_{0.50}$Ca$_{0.50}$MnO$_3$ that this state is in 
fact a prototypical charge density wave (CDW) which undergoes collective transport.
Dramatic resistance hysteresis effects and broadband noise properties are observed, 
both of which are typical of sliding CDW systems. 
Moreover, the high levels of disorder typical of manganites 
result in behaviour similar to that of well-known disordered CDW materials.
Our discovery that the manganite superstructure is a CDW shows that 
unusual transport and structural properties do not require exotic 
physics, but can emerge when a well-understood phase (the CDW) 
coexists with disorder.

The stripe phase in manganites of the form La$_{1-\mathrm{x}}$Ca$_\mathrm{x}$MnO$_3$
appears as the temperature 
is lowered through $T \simeq 240$~K,  and the superstructure 
wavevector settles on a final value of 
\textbf{q}$\simeq(1-x)$\textbf{a$^*$}~\cite{chen_co_melt} 
(where \textbf{a$^*$} is the reciprocal lattice vector) 
for $0.5\leq x < 0.85$, 
at $T \simeq 120$~K~\cite{chen_comm_incomm}.  
Based on the insulating nature of the manganites up to room temperature, 
and the observation of 
stripes of charge order in transmission electron microscopy (TEM) images,
early studies concluded that the superstructure arose from localisation of 
charge at atomic sites~\cite{CO1, CO2a}. 
However, neutron and x-ray studies found the degree of charge localisation at Mn sites to be small,
and subsequent theoretical work suggested that a CDW model may be more applicable~\cite{milward}. 
This suggestion is supported by the observation that $q/a^*$ is strongly temperature 
dependent~\cite{chen_comm_incomm, us}, indicating that a model in which the superstructure 
periodicity is derived from the sample stoichiometry cannot be valid.
In addition, heat capacity peaks at the transition to the stripe phase can be modelled as ``dirty 
Peierls transitions'', expected in a disordered CDW system~\cite{dirty_peierls}.  
However, the possibility of the stripe phase exhibiting sliding behaviour, as
seen in many other CDW systems~\cite{gruner}, could not be probed without the 
ability to make orientation-dependent resistivity measurements.  Here we 
describe the first such measurements on the manganite stripe phase, which reveal 
dramatic orientation-dependent resistivity and broadband noise effects which are
characteristic of CDW sliding.

Orientation-dependent resistivity measurements require
thin films,
because untwinned single crystals of the insulating manganites cannot be grown~\cite{crystal}.
The 80~nm thick \LCMO thin film was grown on an NdGaO$_3$ substrate as described in~\cite{me_strain}.  
The film was prepared for TEM by conventional grinding of the substrate to 50 $\mu$m and then milling a small 
window using a focused ion beam microscope to a thickness of around 200 nm.  The sample was examined in 
a Philips CM30 microscope and was cooled to 90 K using a Gatan liquid nitrogen stage.  
The uniaxial stripe phase was identified via superlattice reflections in a selected area TEM diffraction 
pattern (illustrated below in Figure~\ref{diffres}a); these reflections 
are detectable at 190~K, reaching a stable form at 90~K.  
(Note that previous resistivity measurements of thin film \LCMO have
failed to produce consistent results~\cite{butorin_film, nyeanchi, xiong_film}, 
because of the difficulty of producing high quality films and a
failure to check for the superstructure using a microscopic technique.)

For the resistance measurements, gold wires were attatched to the thin film sample using graphite paint.  
The differential resistivity of the sample studied here was measured
by using a lock-in amplifier to
detect the AC voltage produced in response to a 17~Hz AC current plus a DC bias;
contacts were placed around the edges of the film to enable the
current and bias to be applied along different directions, chiefly
parallel and perpendicular to the superlattice direction.
Both four-point and two-point configurations were employed for the resistivity
experiments to eliminate possible contact resistance effects;
the noise measurements reported below employed two contacts.

\begin{figure}
\begin{centering}
\includegraphics[width=0.95\columnwidth]{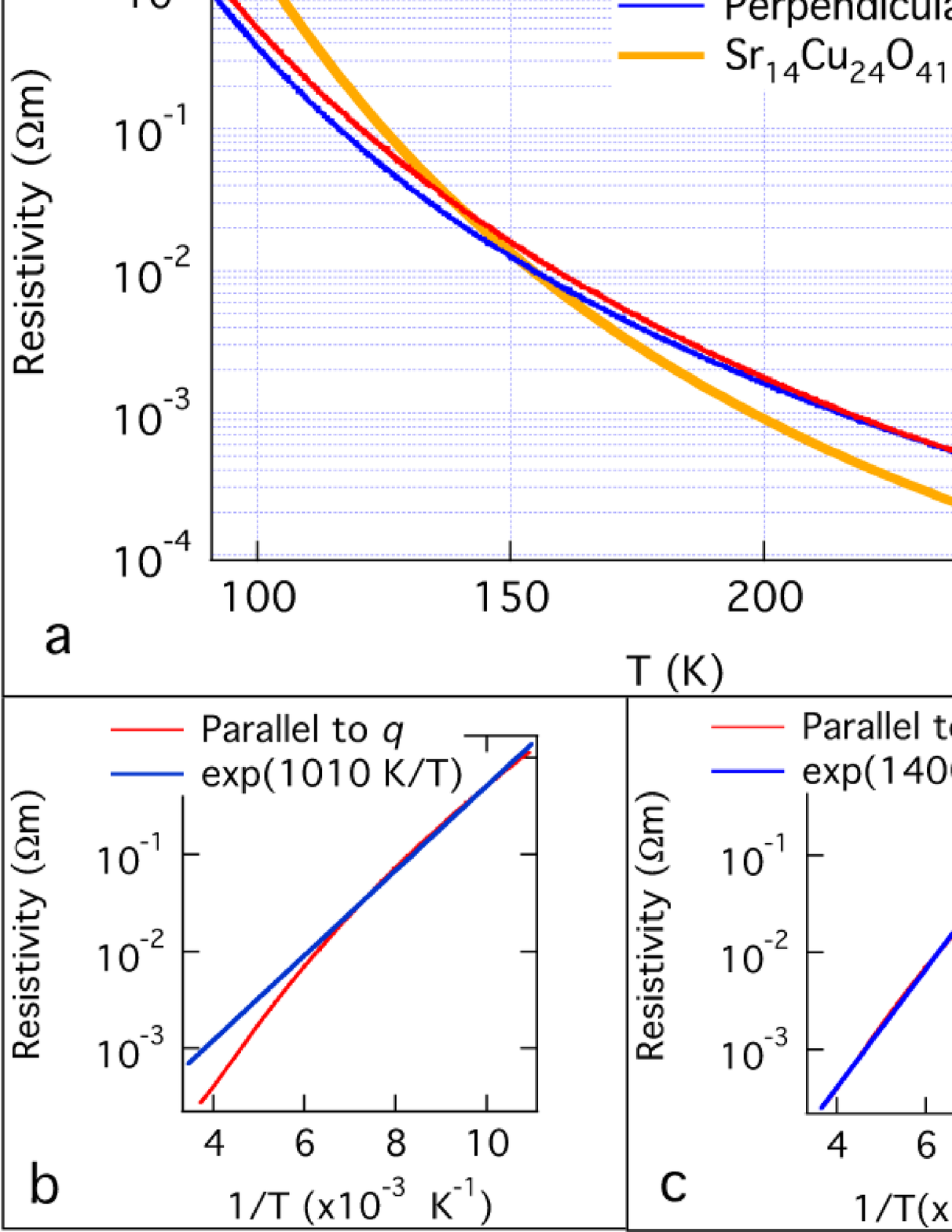}
\caption{\label{R_vs_T}}
\end{centering}
\end{figure}

Analogies between \LCMO and other CDW systems
are clearly apparent in Figure~\ref{R_vs_T}, which shows the 
differential resistivity under zero
DC bias versus temperature.
The measurements appear similar to the resistivities of prototypical CDW 
systems doped with impurities~\cite{pinning_trans_1, pinning_trans_2, pinning_trans_3}, 
in that there is
no clear feature at the expected CDW ordering temperature, with insulating 
behaviour (i.e. decreasing resistivity with increasing temperature) persisting well above it.
This has been interpreted as the ``smearing'' of the transition
caused by the large impurity density~\cite{pinning_trans_1, pinning_trans_2, pinning_trans_3}.  
Analogous behaviour is also seen in cuprate ladder compounds exhibiting sliding 
density waves~\cite{sliding_cuprate_1, sliding_cuprate_2} 
below $T\simeq$200~K. As in the case of the cuprates, 
the resistivity of \LCMO shows an activated temperature dependence
with an activation energy which varies from $\simeq1000-1400$~K (Figure~\ref{R_vs_T}b and c).

Figure~\ref{diffres} shows the differential resistivity as a function of DC bias 
applied parallel (along the lattice vector {\bf a}) and perpendicular to (along the lattice vector {\bf c}) the superlattice direction.
At 157~K (Figure~\ref{diffres}(f)), 
the differential resistivity drops in a similar fashion when
the DC bias is applied in either direction.
However, at temperatures $\ltsim~ 140$~K, the 
differential resistivity undergoes a sharp drop when
the bias is applied in the \textbf{a} direction;
the effect is very much less marked
with the bias in the \textbf{c} direction (Figure~\ref{diffres}(b)-(e)). 
In addition, there is a large hysteresis between the differential
resistance recorded when the bias is
first applied in the \textbf{a} direction after cooling from
300~K, and that measured on subsequent bias sweeps (in Figure~\ref{diffres} b-e the 
upper line in each direction shows the data from the first sweep after cooling and the 
lower line shows the data from subsequent sweeps);
the area enclosed by the hysteresis loop increases rapidly
as the temperature falls.

The hysteretic resistivity features shown in Figure~\ref{diffres} are typical of 
CDWs~\cite{gruner}. As the sample is cooled, the CDW settles into a 
minimum-free-energy pinned configuration, corresponding to maximum electrical resistivity~\cite{gruner}. 
On the application of electric field, 
the CDW initially undergoes local distortions that occur over longer and longer lengthscales as the field increases; 
eventually, the threshold field is reached and the CDW starts to slide. As the field is reduced again, the CDW freezes 
into a distorted state, characterized by a lower resistivity; the initial, minimum energy state cannot be regained without 
thermally cycling the sample~\cite{gruner}, explaining the hysteresis in our data.

Other mechanisms such as avalanche breakdown or sample heating cannot account for the data in Figure~\ref{diffres}. 
Whilst these effects might produce a falling differential resistivity as the field increases, they would 
not produce a history-dependent result; on removing the field, the sample would return to its initial 
state. Moreover, whilst the DC resistivity for currents in the \textbf{c} direction is 2 times higher than that for 
currents parallel to \textbf{a}, the drop in resistivity as the field increases is five times larger in the latter 
direction (Figure~\ref{diffres}); this anisotropy both fits naturally into the CDW picture and excludes heating 
and breakdown as possible mechanisms. The anisotropy in the observed
effects also excludes ferromagnetic domains (sample inhomogeneity) as a possible mechanism;
in this case the effects would be the same in the two orientations.

Having explained the hysteresis when the bias is along {\bf a}, we attribute
the small amount of hysteresis seen when the bias is along \textbf{c}
(Figure~\ref{diffres}) to imperfect 
contact geometry; i.e. misalignment results in a small amount of bias being
applied in the perpendicular direction.

Another distinguishing feature of CDWs is that they exhibit
a broadband noise spectrum with an amplitude proportional to
$f^{-\alpha}$, where $f$ is the 
frequency~\cite{Maeda1983, Maeda1985,  Richard, maher,  zettl&gruner}.
Noise measurements were performed using a low-noise current source.  The noise signal was amplified with 
a high input impedence, low noise preamplifier and was recorded via a digitizing oscilloscope.
Lead capacitance, and the typical $10^6~\Omega$ sample resistance, limited noise measurements to below 
10~kHz.  Other techniques commonly used on CDW systems such as NbSe$_3$ were considered
or attempted but typical properties of manganite films rendered them
impossible; manganite film resistivities and geometries lead to RC time
constants that are too high to perform experiments that measure effective
pulse line or duration memory effects.

Figure~\ref{noise} shows that significant broadband noise is observed in \LCMO when 
the DC bias is applied in the superstructure direction.
By contrast, the noise amplitude is much smaller with the bias in 
the non-superstructure direction.
The exponent $\alpha$ in \LCMO runs from 0.8 (156~K) to 2.0 (100~K),
a similar range to values seen in the prototypical CDW system NbSe$_3$ (0.8-1.8)~\cite{gruner}. 
However, the magnitude of the broadband noise in \LCMO
is much larger than that 
observed in clean CDW systems; for \LCMO the effective noise temperature at 300~Hz
is $\sim10^{11}$~K for a sample temperature of 100~K, while in pure NbSe$_3$ the 
effective noise temperature is $\sim10^6$~K.  
This is attributable to the large amount of disorder 
present in \LCMO~\cite{dirty_peierls}
(see above), 
so that there are many more pinning-depinning events
compared to e.g. pure NbSe$_3$.
Although broadband noise has previously been observed in impurity-pinned 
CDWs~\cite{BBN_imps}, narrowband noise has not been observed
in an impurity-doped or radiation-damaged sample, probably because
the width of the narrowband noise peak is proportional to the 
magnitude of the broadband noise~\cite{maher}.
Therefore a high level of disorder or impurity pinning will lead to a large amount of 
broadband noise and unobservably small narrowband noise, as seen here in \LCMO.

As seen in other CDW systems~\cite{zettl&gruner},
the amount of broadband noise decreases with 
increasing temperature (Figures~\ref{noise}a,c,e).
For a bias above $E_\mathrm{T}$($\simeq10^4$V/m) in the superstructure direction
this decrease is approximately linear with temperature (Figure~\ref{noise_graphs}c),
as observed in the CDW system TaS$_3$~\cite{zettl&gruner}.
With the bias in the non-superstructure direction,
the noise increases much more slowly; at 100~K it is more than an order of magnitude
smaller than that with the bias along {\bf a} (Figure~\ref{noise_graphs}c).

Figures~\ref{noise_graphs}a and b show the variation of the broadband noise 
amplitude with applied bias between the first bias sweep 
after cooling from 300~K and on a subsequent sweep.
On first biasing, the noise ampitude shows a large peak at the
same point as the large fall in differential 
resistance (Figure~\ref{noise_graphs}a).  
On subsequent bias sweeps, the noise
increases more slowly with bias (Figure~\ref{noise_graphs}b).
The large peak during the first bias sweep is caused by a high level of 
random telegraph signal (RTS) noise, which occurs in CDW 
systems as they switch from pinned to depinned states and back 
again~\cite{RTS_1, RTS_2} close to the threshold field.
The distinctive shape of the RTS noise is shown in Figure~\ref{noise_graphs}d, another
factor adding weight to our identification of a CDW in \LCMO.

In conclusion, we have demonstrated via resistivity and noise 
measurements that the superstructure
in the stripe phase of manganites is a CDW which slides in 
the presence of an electric
field. The manganite CDW is a fully gapped system with no screening electrons,
which has previously only been observed in extremely clean organic 
materials~\cite{Ross}. However, the
manganite CDW exists with a high level of impurities, leading to 
dramatic hysteresis effects in
the resistance. Our findings call for a reanalysis of the large 
region of the manganite phase diagram, $0.5 \leq x < 0.85$,
which is occupied by the CDW.
This is the first observation of sliding in a material which
undergoes 3D charge ordering~\cite{Radaelli}.
In a wider context, this result is important because of the 
prevalence of stripe and checkerboard phases
in oxide materials, including chelates, cobaltites, nickelates and 
cuprates; in particular, evidence is
mounting that a glass of the stripe phase in cuprates is linked to 
high temperature superconductivity~\cite{cup_glass}, making an understanding of the stripe phase a matter of urgency.

We thank N. Harrison, N.D. Mathur, P.A. Midgley, G. Kotliar and E. Rosten for helpful comments.  S.~Cox acknowledges support
from the Seaborg institute.  
The sample was grown at Cambridge where research was funded by the UK EPSRC. 
This research was funded by the U.~S. Department of Energy (DoE) under 
Grant LDRD-DR 20070013. Work at NHMFL is performed under the auspices of the
NSF, the State of Florida, and the US DoE.

The authors declare no competing financial interests.

\newpage
\noindent Figure 1: (a) Differential resistance of \LCMO with the current
in the \textbf{a} (red) and \textbf{c} (blue) directions versus temperature (zero DC bias).
The resistivity is similar to that of the ladder compound Sr$_{14}$Cu$_{24}$O$_{41}$,
shown in yellow~\cite{sliding_cuprate_1}. 
(b) and (c) demonstrate that the resistivity is activated over all temperatures,
being fitted to two exponentials.

\vspace{0.45in}
\noindent Figure 2: (a) Linescan of TEM image in \textbf{a} (red)
and \textbf{c} (blue) directions showing the superstructure
reflections present in only the \textbf{a} direction.
(b)-(f)~Differential resistivity of \LCMO versus DC bias
with bias applied
in the \textbf{a} (red) and \textbf{c} (blue) directions at various temperatures.
In each case the upper curve is the differential 
resistivity obtained after cycling the temperature to 300~K, and the lower curve is the 
path followed by subsequent bias sweeps.

\vspace{0.45in}
\noindent Figure 3: Frequency and current depedence of the broadband noise.  97~K data with the current 
parallel (a) and perpendicular (b) to $q$.  123~K data with the current parallel (c) and perpendicular (d) 
to $q$.  156~K data with the current parallel (e) and perpendicular (f) to $q$.  The color scale is in units 
of V$^2$/Hz.

\vspace{0.45in}
\noindent Figure 4: (a) Resistivity displayed as $R(E)/R(0)$ (red), and noise signal at 300~Hz (blue) for the first time current is passed parallel to the 
superstructure after cooling from 300~K.   (b) Resistivity (red) and noise (blue) for the second time current is passed.
(c) Noise signal at 300~Hz and 10~V as a function of temperature in the superstructure (red) 
and non-superstructure (blue) directions. The noise at 300~Hz was extracted by calculating the power spectral 
density of the noise and then taking the square root. Twenty points of the power spectral density were averaged.
(d)Noise signal a short amount of time after the current has been changed; first (red) and second (blue)
time the current is swept.

\newpage

\begin{figure}
\begin{centering}
\includegraphics[width=0.9\columnwidth]{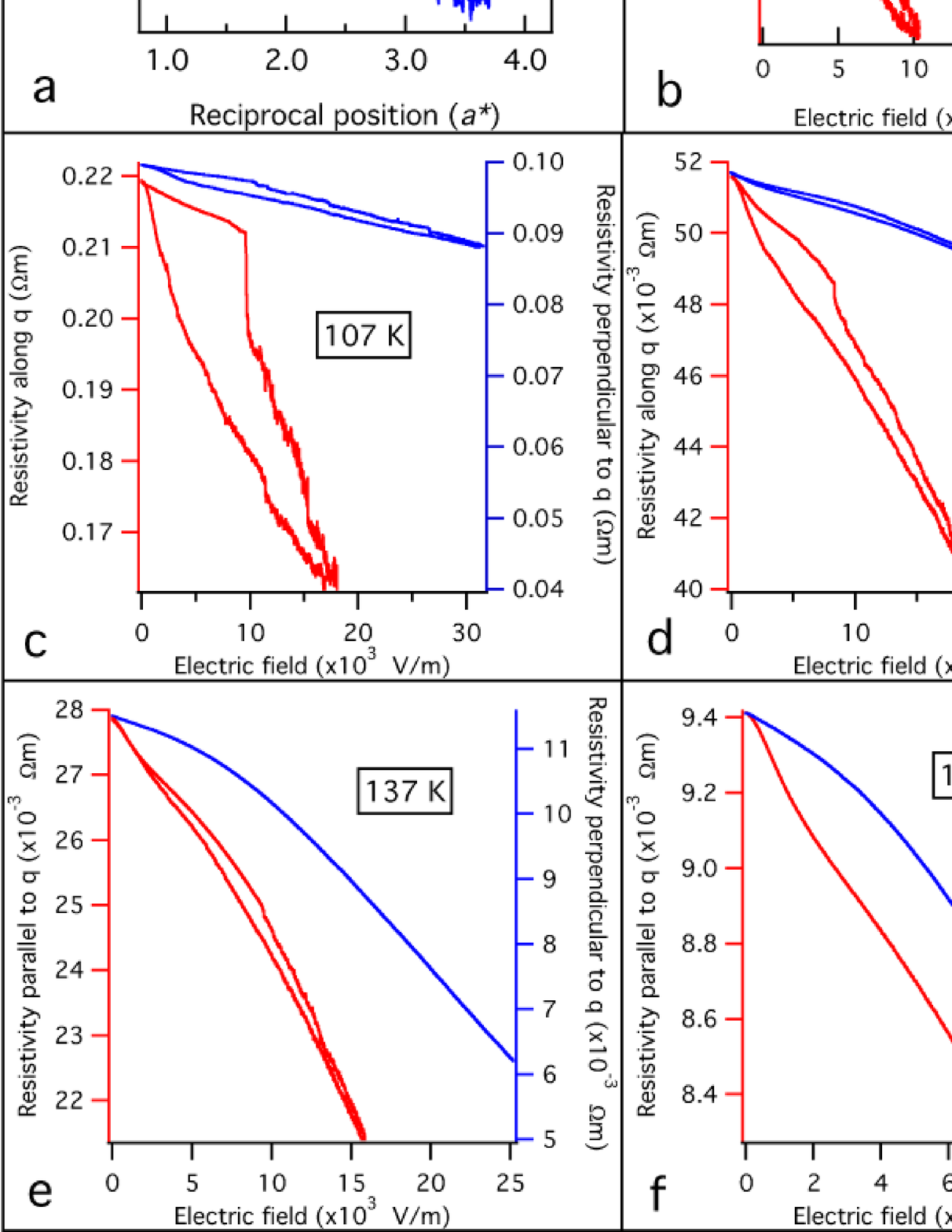}
\caption{\label{diffres}}
\end{centering}
\end{figure}

\begin{figure}
\begin{centering}
\includegraphics[width=0.98\columnwidth]{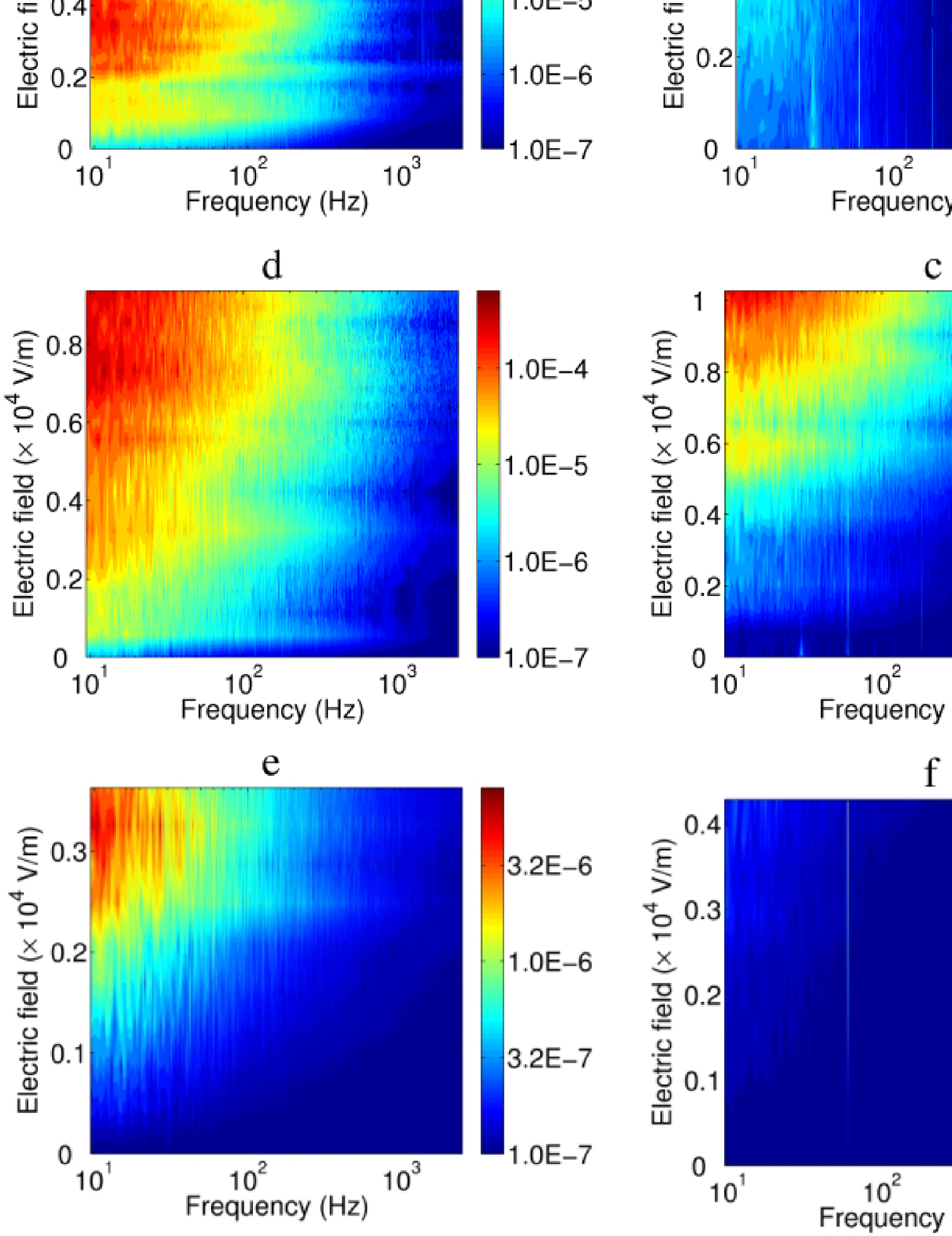}
\caption{\label{noise}}
\end{centering}
\end{figure}

\begin{figure}
\begin{centering}
\includegraphics[width=0.95\columnwidth]{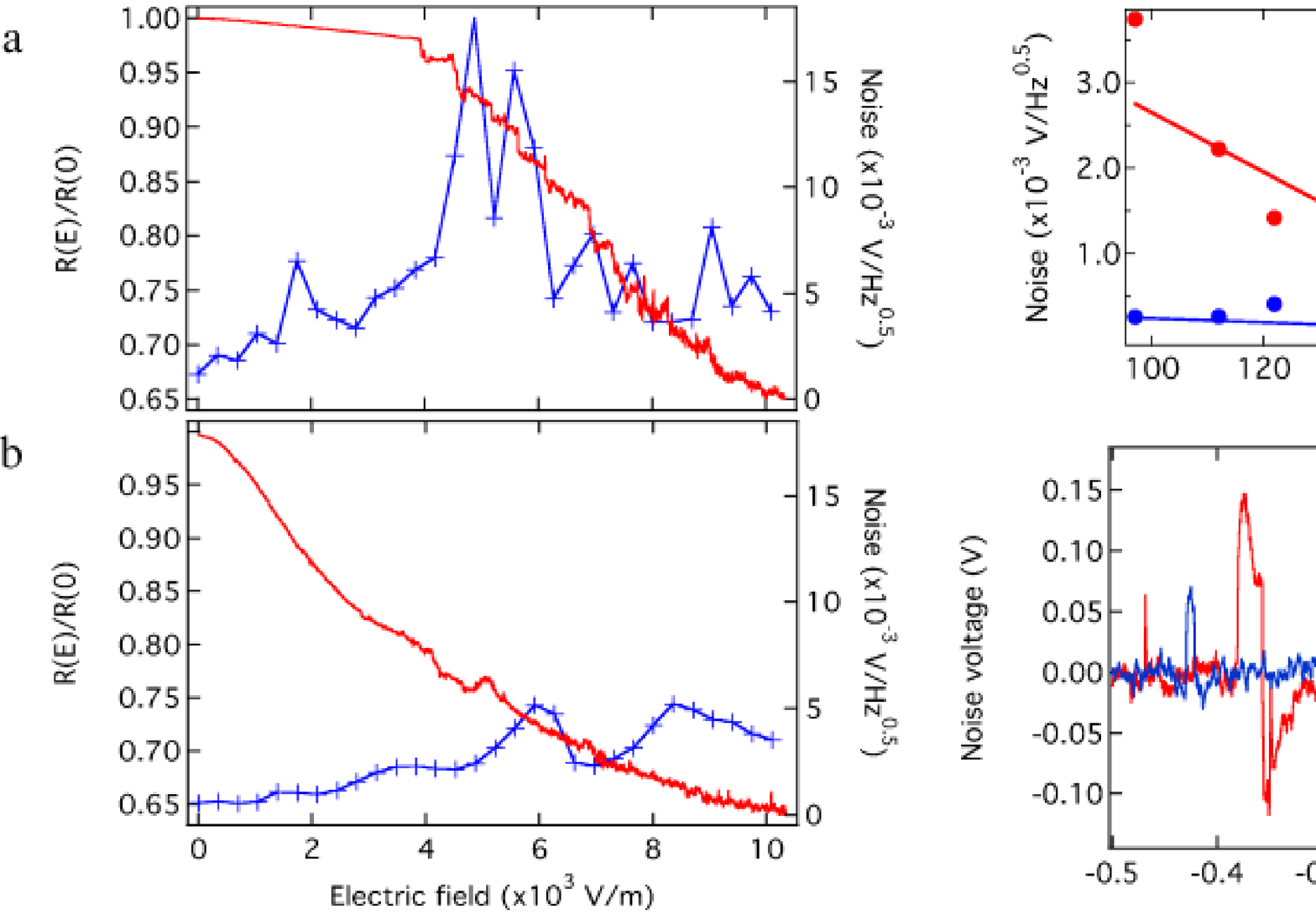}
\caption{\label{noise_graphs}}
\end{centering}
\end{figure}

\newpage


\begin{thebibliography}{10}

\bibitem{chen_co_melt}
{Chen, C. H., Mori, S. \& Cheong, S.-W.}
\newblock Anomalous melting transition of the charge-ordered state in
  manganites.
\newblock {\em Phys. Rev. Lett.}{\bf, 83}, 4792--4795, 1999.

\bibitem{chen_comm_incomm}
{Chen, C. H. \& Cheong, S.-W.}
\newblock Commensurate to incommensurate charge ordering and its real-space
  images in {La$_{0.5}$Ca$_{0.5}$MnO$_3$}.
\newblock {\em Phys. Rev. Lett.}{\bf, 76}, 4042--4045, 1996.

\bibitem{CO1}
{Chen, C. H., Cheong, S.-W. \& Hwang, H.Y.}
\newblock Charge-ordered stripes in {La$_{1-x}$Ca$_x$MnO$_3$ with x$>$0.5}.
\newblock {\em J. Appl Phys}{\bf, 81}, 4326, 1997.

\bibitem{CO2a}
{Mori, S., Chen, C. H. \& Cheong, S.-W.}
\newblock Pairing of the charge-ordered stripes in {(La,Ca)MnO$_3$}.
\newblock {\em Nature}{\bf, 392}, 473--476, 1998.

\bibitem{milward}
{Milward, G.C., Calder\'{o}n, M.J., \& Littlewood, P.B}.
\newblock Electronically soft phases in manganites.
\newblock {\em Nature}{\bf, 433}, 607--610, 2005.

\bibitem{us}
{Loudon, J.C. \emph{et al.}}
\newblock Weak charge-lattice coupling requires reinterpretation of stripes of
  charge order in {La$_{1-x}$Ca$_x$MnO$_3$}.
\newblock {\em Phys. Rev. Lett.}{\bf, 94}, 097202/1--4, 2005.

\bibitem{dirty_peierls}
{Cox, S.~\emph{et al.}}
\newblock Evidence for the charge-density-wave nature of the stripe phase in
  manganites.
\newblock {\em J. Phys.: Condens. Matter}{\bf, 19}, 192201/1--7, 2007.

\bibitem{gruner}
G.~Gr\"{u}ner.
\newblock {\em Density waves in solids}.
\newblock Addison-Wesley, 1994.

\bibitem{crystal}
{Majewski, P., Epple, L., Rozumek, M. \& Schluckwerder, H.}
\newblock Phase diagram studies in the quasi binary systems
  {LaMnO$_3$--SrMnO$_3$ and LaMnO$_3$--CaMnO$_3$}.
\newblock {\em J. Mater. Res.}{\bf, 15}, 1161--1166, 2000.

\bibitem{me_strain}
{Cox, S.~\emph{et al.}}
\newblock Strain control of superlattice implies weak charge-lattice coupling
  in {La$_{0.5}$Ca$_{0.5}$MnO$_3$}.
\newblock {\em Phys. Rev. B}{\bf, 73}, 132401/1--4, 2006.

\bibitem{butorin_film}
{Butorin, S.M., S\aa the, C., Saalem. F., Nordgren, J. \& Zhu, X.M.}
\newblock {Probing the Mn$^{3+}$ sublattice in La$_{0.5}$Ca$_{0.5}$MnO$_3$ by
  resonant inelastic soft S-ray scattering at the Mn L$_{2,3}$ edge}.
\newblock {\em Surface Review and Letters}{\bf, 9}, 989--982, 2002.

\bibitem{nyeanchi}
{Nyeanchi, E.B., Krylov, I.P., Zhu, X.-M. \& Jacobs, N.}
\newblock Ferromagnetic ground state in {La$_{0.5}$Ca$_{0.5}$MnO$_3$} thin
  films.
\newblock {\em Euro. Phys. Lett.}{\bf, 48}, 228--232, 1999.

\bibitem{xiong_film}
{Xiong, Y.M. \emph{et al.}}
\newblock Magnetotransport properties in {La$_{1-x}$Ca$_x$MnO$_3$} (x=0.33,0.5)
  thin films deposited on different substrates.
\newblock {\em J. Appl. Phys.}{\bf, 97}, 083909/1--11, 2005.

\bibitem{pinning_trans_1}
{Ong, N.P.~\emph{et al.}}
\newblock Effect of impurities on the anomalous transport properties of
  {NbSe$_3$}.
\newblock {\em Phys. Rev. Lett.}{\bf, 42}, 811--814, 1979.

\bibitem{pinning_trans_2}
{Brill, J.W.~\emph{et al.}}
\newblock Impurity effect on the frolich conductivity in {NbSe$_3$}.
\newblock {\em Phys. Rev. B}{\bf, 23}, 1517--1526, 1981.

\bibitem{pinning_trans_3}
{Chaikin, P.M.~\emph{et al.}}
\newblock Thermopower of doped and damaged {NbSe$_3$}.
\newblock {\em Sol. Stat. Comm.}{\bf, 39}, 553--557, 1981.

\bibitem{sliding_cuprate_1}
{Blumberg, G.~\emph{et al.}}
\newblock Sliding density wave in {Sr$_{14}$Cu$_{24}$O$_{41}$} ladder
  compounds.
\newblock {\em Science}{\bf, 297}, 584--587, 2002.

\bibitem{sliding_cuprate_2}
{Maeda, A.~\emph{et al.}}
\newblock Sliding conduction by the quasi-one-dimensional charge-ordered state
  in {Sr$_{14-x}$Ca$_x$Cu$_{24}$O$_{41}$}.
\newblock {\em Phys. Rev. B}{\bf, 67}, 115115/1--5, 2003.

\bibitem{Maeda1983}
{Maeda, A., Naito, M. \& Tanaka, S.}
\newblock Broad and narrow band noise of monoclinic {TaS$_3$}.
\newblock {\em Sol. Stat. Comm.}{\bf, 47}, 1001--1005, 1983.

\bibitem{Maeda1985}
{Maeda, A., Naito, M. \& Tanaka, S.}
\newblock Nonlinear conductivity and broad band noise of monoclinic {TaS$_3$}.
\newblock {\em J. Phys. Soc. Jpn.}{\bf, 54}, 1912--1922, 1985.

\bibitem{Richard}
{Richard, J., Monceau, P., Papoular, M. \& Renard, M.}
\newblock {f$^{-\alpha}$} noise in NbSe$_3$.
\newblock {\em J. Phys. C}{\bf, 15}, 7157--7164, 1982.

\bibitem{maher}
{Maher, M.P.~\emph{et al.}}
\newblock Size effects, phase slip, and the origin of {f$^{-\alpha}$ noise in
  NbSe$_3$}.
\newblock {\em Phys. Rev. B}{\bf, 43}, R9968--9971, 1991.

\bibitem{zettl&gruner}
{Zettl, A. \& Gr\"{u}ner, G.}
\newblock Broad band noise associated with the current carrying charge density
  wave state in {TaS$_3$}.
\newblock {\em Sol. Stat. Comm.}{\bf, 46}, 29--32, 1983.

\bibitem{BBN_imps}
{Maeda, A., Uchinokura, K. \& Tanaka, S.}
\newblock The effect of strong impurities on the low-frequency broad-band noise
  in {NbSe$_3$}.
\newblock {\em Synthetic Metals}{\bf, 19}, 825--830, 1987.

\bibitem{RTS_1}
{Marley, A.C., Bloom, I. \& Weissman, M.B.}
\newblock Temperature and electric-field dependences of characteristic noise
  patterns in mesoscopic ortho-{TaS$_3$} charge density waves.
\newblock {\em Phys. Rev. B}{\bf, 49}, 16156--16161, 1994.

\bibitem{RTS_2}
{Bloom, I., Marley, A.C. \& Weissman, M.B.}
\newblock Discrete fluctuators and broad-band noise in the charge-density-wave
  in {NbSe$_3$}.
\newblock {\em Phys. Rev. B}{\bf, 50}, 5081--5088, 1994.

\bibitem{Ross}
{McDonald, R.D.~\emph{et al.}}
\newblock Charge-density waves survive the pauli paramagnetic limit.
\newblock {\em Phys. Rev. Lett.}{\bf, 93}, 076405/1--4, 2004.

\bibitem{Radaelli}
{Radaelli, P.G., Cox, D., Marezio, M., Cheong, S.-W, Schiffer, P.E. \& Ramirez,
  A.P.}
\newblock Simultaneous structural, magnetic, and electronic transtions in
  la$_{1-x}$ca$_x$mno$_3$ with $x$=0.25 and 0.5.
\newblock {\em Phys. Rev. Lett.}{\bf, 75}, 4488, 1995.

\bibitem{cup_glass}
{Kohsaka, Y.~\emph{et al.}}
\newblock An intrinsic bond-centered electronic glass with unidirectional
  domains in underdoped cuprates.
\newblock {\em Science}{\bf, 315}, 1380--1385, 2007.

\end{thebibliography}
\end{document}